\documentclass[%
  reprint,
 superscriptaddress,
 amsmath,amssymb,
 aps,UTF8,
]{revtex4-1}

\usepackage{graphicx}
\usepackage{dcolumn}
\usepackage{bm}
\usepackage{xcolor}
\usepackage{url}
\usepackage{float}
\usepackage{array}

\usepackage{booktabs}

\usepackage{subfigure}
\usepackage{multirow}

\begin{document}


\title{Novel Neutron Star Structures with Nucleon Mass Radius}
\author{Wei Kou}
\email{kouwei@impcas.ac.cn}
\affiliation{Institute of Modern Physics, Chinese Academy of Sciences, Lanzhou 730000, China}
\affiliation{University of Chinese Academy of Sciences, Beijing 100049, China}
\author{Xiaopeng Wang}
\email{wangxiaopeng@impcas.ac.cn}
\affiliation{Institute of Modern Physics, Chinese Academy of Sciences, Lanzhou 730000, China}
\affiliation{University of Chinese Academy of Sciences, Beijing 100049, China}
\affiliation{Lanzhou University, Lanzhou 730000, China}

\author{Xurong Chen}
\email{Corresponding author: xchen@impcas.ac.cn}
\affiliation{Institute of Modern Physics, Chinese Academy of Sciences, Lanzhou 730000, China}
\affiliation{University of Chinese Academy of Sciences, Beijing 100049, China}
\affiliation{Guangdong Provincial Key Laboratory of Nuclear Science, Institute of Quantum Matter, South China Normal University, Guangzhou 510006, China}



\begin{abstract}
In order to reveal the difference between the latest neutron star observation experiment GW170817 and the existing theory, we mainly consider the effect of the nucleon radius on the neutron star from the existing theory. We believe that the effect of nucleon radius in neutron star is not negligible, and the mass radius of nucleon should be used instead of the charge radius. The nucleon mass radius is set as $r_m = 0.55\pm0.09$ fm from the new measurements. It is considered as an input to the ``Excluded Volume Effects" model in the equation of state of nuclear matter. We propose a novel neutron star mass-radius relation by using proton mass radius is consistent with the observation GW170817.

\end{abstract}

\maketitle


\section{Introduction}
\label{sec:intro}
The properties of neutron stars (NS) are determined by their internal components. The formation of NS is a hot spot problem in which the nuclear physics, particle physics and astrophysics and their combinations. The discovery of GW170817 \cite{Capano:2019eae}, a pair of NS, raises the climax of the research. In general, NS are the densest observable objects in the universe, and they are excellent natural laboratories for studying the state of matter under extreme conditions in nuclear physics and Quantum Chromodynamics (QCD). As a key input to determine the structure of a NS, the equation of state (EOS) of dense nuclear matter is particularly important. 

  EOS of nuclear matter is a good means to explore the structures and properties of NS. A number of models have been developed in recent decades to explain the astrophysical observations \cite{Ozel:2016oaf}. Most models are based on Thomas-Fermi model (See Ref. \cite{Shen:2011qu}) or Relativistic Mean Field (RMF) model \cite{Serot:1984ey} and Hartree-Fock model \cite{Machleidt:1989tm}. To obtain the EOS of the nuclear matter, the RMF model is based on the single boson exchange mechanism of the nuclear force, and the relativistic Lagrangian that satisfies the basic symmetry is constructed. The meson fields are approximated as the mean fields in the multi-body problems, and the ground state and excited state properties of medium-mass nuclei are described successfully (See the Sec. \ref{subsec:rmf} for more).
 
 As a typical application of RMF, the NL3 \cite{Lalazissis:1996rd} parameters set gives a good description of the EOS of nuclear matter, but it has a slight deviation when calculating the structures and properties of NS. The prediction for a NS radius by NL3 which have 1.4 solar masses is about 14-15 km. But in the most recent experimental observation, GW170817, the radius of a NS with the same mass was measured which is about $R_{1.4M_\odot}=11.0_{-0.6}^{+0.9}$ km (90$\%$ credible interval) \cite{Capano:2019eae}. In the Ref. \cite{Fattoyev:2010mx,Mart:2013gfa,Suparti:2017msx}, the authors used a new parameters set, IUFSU, and considered the effects of the nucleon radius on the structure of the NS. Their results are close to the latest observations. But the past two years ago, other authors \cite{Zhu:2018qpv,Zhu:2018vwn} have used the quark mean field bag model to calculate the NS radius on the basis of previous studies. The latter thought that the nucleon radius had little effect on the structures of NS. There are many models and experiments for NS radius predictions. For more information, please refer to the review article \cite{Ozel:2016oaf}.
 
 Nucleons radii are the elementary physical quantity of themselves. They have been measured for a long history. The current data about the ``Charge" radius of the proton is fixed at $r_{ch} = 0.8409\pm0.0004$ fm and ``Magnetic" radius of the neutron at $r_{mag} = 0.864\pm0.008$ fm \cite{ParticleDataGroup:2020ssz}. In previous work we discussed in last paragraph, the authors introduced the charge or magnetic radii of the nucleons into EOS and obtained different results. But in dense nuclear matter, the nucleons are very close together at a narrow space. The strong interaction is dominant and the ``Range" of the medium particles transmitting the strong interaction is smaller than the nucleon charge radius. Neutron stars, on the other hand, are dominated by neutrons. Neutrons are electrically neutral and therefore do not participate in electromagnetic interactions. The authors of Ref. \cite{Zhu:2018qpv} believe that the viewpoint in Ref. \cite{Mart:2013gfa,Suparti:2017msx} is not accurate enough, and they believe that the free radius of nucleons has little influence on the calculation results of the whole NS. In the latter literature \cite{Suparti:2017msx}, RMF parameters were re-fitted with different nucleons charge radii. 
 
 As stated in the previous discussions, the effects of nucleons charge and magnetic radii are small in dense nuclear matter, which are consistent with the conclusions in Ref. \cite{Zhu:2018qpv}. But they are not all stories. In recent years, the problems of the origin of proton mass has attracted much attention \cite{Ji:2021qgo,Ji:2021pys,Ji:2021mtz,Kharzeev:2021qkd,Wang:2021dis,Kou:2021bez,Kou:2021qdc}. The mass radius of the proton is introduced into the researches of QCD. In our previous work \cite{Wang:2021dis}, we used the vector meson photoproduction data near the threshold \cite{Barth:2003kv,LEPS:2005hax,GlueX:2019mkq} to obtain the gravitational form factor and the mass radius of proton. It is about $r_m= 0.55\pm0.09$ fm and obviously smaller than charge one. On this basis, we believe that the nucleon radius has an influence on the structure of the NS, but it should be dominated by the mass radius of the nucleon rather than the charge one.

This paper is divided into the following parts: In Sec. \ref{sec:form}, we mainly introduced the idea of RMF to calculate the EOS of nuclear matter, and adopted excluded volume effects (EVE) model \cite{Rischke:1991ke,Kouno:1995fs} to add the mass radius of nucleons from Ref. \cite{Kharzeev:2021qkd,Wang:2021dis} as the correction. At the end of Sec. \ref{sec:form}, we briefly explained how to get the radius and mass information of NS from EOS. In the Sec. \ref{sec:Results and discussions} we present the results of our work and the discussions. The last section is the summary.

\section{FORMALISM}
\label{sec:form}
\subsection{Relativistic Mean Field Theory}
\label{subsec:rmf}
Using the relativistic mean field (RMF) theory \cite{Serot:1984ey} has worked well in describing the EOS of NS. To carry out the study of nuclear matter, the meson-exchange model should be introduced in standard Lagrangian to describe the nucleon-nucleon interaction. The starting point of RMF theory is a standard Lagrangian density \cite{Gambhir:1989mp,Lalazissis:1996rd}
\begin{equation}
	\begin{aligned}
		\mathcal{L}=& \bar{\psi}\left(\gamma\left(i \partial-g_{\omega} \omega-g_{\rho} \vec{\rho} \vec{\tau}-e A\right)-m-g_{\sigma} \sigma\right) \psi\\
		&+\frac{1}{2}(\partial \sigma)^{2} 
		-U(\sigma)-\frac{1}{4} \Omega_{\mu \nu} \Omega^{\mu \nu}+\frac{1}{2} m_{\omega}^{2} \omega^{2}\\
		&-\frac{1}{4} \overrightarrow{\mathrm{R}}_{\mu \nu} \overrightarrow{\mathrm{R}}^{\mu \nu}+\frac{1}{2} m_{\rho}^{2} \vec{\rho}^{2} \\
		&-\frac{1}{4} \mathrm{~F}_{\mu \nu} \mathrm{F}^{\mu \nu},
	\end{aligned}
\label{eq:lagrangian}
\end{equation}
which contains nucleons mass $m$; scalar and vector mesons $\sigma$, $\omega$ and $\rho$; the electromagnetic field; and nonlinear self-interactions of
the scalar $\sigma$ field,
\begin{equation}
	U(\sigma)=\frac{1}{2} m_{\sigma}^{2} \sigma^{2}+\frac{1}{3} g_{2} \sigma^{3}+\frac{1}{4} g_{3} \sigma^{4}.
	\label{eq:potential}
\end{equation}
The total antisymmetric tensors of the vector mesons can be written as
\begin{equation}
	\begin{aligned}
		\Omega_{\mu \nu} &=\partial_{\mu} \omega_{\nu}-\partial_{\nu} \omega_{\mu}, \\
	\overrightarrow{\mathrm{R}}_{\mu \nu} &=\partial_{\mu} \rho_{\nu}-\partial_{\nu} \rho_{\mu}.
	\end{aligned}
	\label{eq:Tensor}
\end{equation}

Generally speaking, each parameter in Lagrangian is given by fitting the experimental data of nuclear physics. Over the last few decades, physicists have used some of the following parameter sets most frequently: NL1 \cite{Reinhard:1986qq}, NL3 \cite{Lalazissis:1996rd}, NL-SH \cite{Sharma:1993it}, PL-40 \cite{Sharma:1991ze}, FSU \cite{ToddRutel:2005fa} and its update version IU-FSU \cite{Fattoyev:2010mx}. The last one considered the vector mesons self-interactions and $\rho-\omega$ couplings, so it has more parameters to describe the interactions in nuclear matter system. Since our focus is on how much the radius of the nucleons corrects the EOS of nuclear matter, in this work we only consider simple scalar meson self-interactions by using the NL3 parameters set. In NL3 parameters set, we use $m = 939$ MeV, $m_\sigma = 508.194$ MeV, $m_\omega=782.501$ MeV and $m_\rho = 763$ MeV. Table \ref{tab:NL3} represents the fundamental parameters by NL3 model \cite{Lalazissis:1996rd} which describe the coupling constant of meson-exchange interactions.

\begin{table}[H]
	\caption{Parameter sets for the NL3 models discussed in the text.}
	\begin{center}
		\begin{tabular}{ c|c|c|c|c }
			\toprule[1.3pt]
			  $g_2$ (fm$^{-1}$)& $g_3$ (MeV)& $g_\sigma$ (MeV)& $g_\omega$ (MeV)& $g_\rho$ (MeV)\\
			\hline
			   $-10.431$ & $-28.885$ & $10.217$ & $12.868$ &$4.474$\\
			
			\bottomrule[1.3pt]
		\end{tabular}	
		\label{tab:NL3}
	\end{center}
\end{table} 

Based on the previous discussions, we easily obtain the energy density of the nuclear matter system by Legendre transformation of the Lagrangian. In the same ways we can also get the filed equations of nucleons and mesons by using the Euler-Lagrange equation. We will discuss them in Sec. \ref{subsec:eos}.

\subsection{EOS by Correction from Nucleons Mass Radii}
\label{subsec:eos}
Before calculating the EOS of nuclear matter, we first consider the thermodynamic properties of nuclear matter. In multi-body physics, of course, we cannot solve the motion states of every particle. Therefore, in the multi-body nuclear matter, a large number of nucleons are treated as approximate the ideal Fermi gases. One can use thermodynamic quantities to describe the properties of nucleons behaviors, such as chemical potential and pressure and so on. In this work, we are inspired by previous researches \cite{Rischke:1991ke,Kouno:1995fs,Mart:2013gfa,Suparti:2017msx,Zhu:2018qpv,Zhu:2018vwn} and we also think that the radii of the nucleons have effects on the structure of the NS -- mass-radius relations.

The effect of nucleon volume on the EOS of nuclear matter is introduced by Ref. \cite{Rischke:1991ke,Kouno:1995fs}. In the mean field theory for point-like nucleons, at zero temperature, the baryon density $\rho_B$ is given as
\begin{equation}
	\rho_{B}=\frac{p_{F}^{3}}{3 \pi^{2}}, 
	\label{eq:density}
\end{equation}
where $p_{F}$ is the Fermi momenta of nucleon $B$ and $B = n,\ p$. In general case the nuclear matter has $N$ nucleons and a volume $V$ or $\rho=N/V$. If we consider the volume effect of the nucleons then we must make a correction for the density. In the EVE model \cite{Rischke:1991ke,Kouno:1995fs}, the volume of nuclear matter $V$ for a system of $N$ nucleons in configurational space is
reduced to an effective one, $V^{\prime}=V-N V_{n}$, where $V_n=4\pi r_m^3/3$ is the volume of a nucleon and $r_m$ is the nucleon mass radius which can be set as $0.55 \pm 0.09$ fm \cite{Kharzeev:2021qkd,Wang:2021dis}. In other words, nucleons of volume $V_n$ in volume $V$ are described as a ``point-like particle" and are moving thermally in an effective volume $V^{\prime}$. In this case, we can redefine the particle density as $\rho^{\prime}=N / V^{\prime}$ which is equal to the density $\rho_{B}$ for the given $k_F$. In this way, the true nucleon density in nuclear matter is written as follows:
\begin{equation}
	\rho=\frac{\rho^{\prime}}{1+V_{n} \rho^{\prime}}.
	\label{eq:bayon density}
\end{equation}
Because of the consistency of thermodynamic quantities, the scalar density of the nucleons in the whole system can also be written as
\begin{equation}
	\rho_{s}=\frac{\rho_{s}^{\prime}}{1+V_{n} \rho^{\prime}},
	\label{eq:rhoscalar}
\end{equation}
where
\begin{equation}
	\begin{aligned}
		\rho_{s}^\prime &=\frac{1}{\pi^{2}} \sum_{i=n, p} \int_{0}^{p_{F}^{i}} d p p_{i}^{2} \frac{M_{N}^{*}}{\sqrt{M_{N}^{* 2}+p_{i}^{2}}} \\
		&=\frac{M_{N}^{*}}{2 \pi^{2}}\left(p_{F}^{i} E_{F}^{i}-M_{N}^{* 2} \ln \left|\frac{p_{F}^{i}+E_{F}^{i}}{M_{N}^{*}}\right|\right),\\
		E_{F}^{i}&=\sqrt{M_{N}^{* 2}+\left(p_{F}^{i}\right)^{2}},
		\label{eq:rhoscalarINT}
	\end{aligned}
\end{equation}
and the effective mass $M_N^* = m+g_\sigma \sigma$ for nucleons come from Eq. (\ref{eq:lagrangian}), $p_F^n(p_F^p)$ come from Eq. (\ref{eq:density}) but have the corrections with EVE model \cite{Rischke:1991ke,Kouno:1995fs}. 

In the Lagrangian density corresponding to nuclear matter, the meson fields exist in the form of quantum fields. Because it is very difficult to calculate the quantum fields of multi-body, the mean field approximation method is needed to replace the meson fields with the mean fields and change the field equations into an algebraic equations to solve easily. Applying the Euler-Lagrange equation
\begin{equation}
	\frac{\partial \mathcal{L}}{\partial \phi}-\partial^{\mu}\left[\frac{\partial \mathcal{L}}{\partial\left(\partial^{\mu} \phi\right)}\right]=0,
	\label{eq:E-L equation}
\end{equation}
we can then take the different fields and substitute them into Eq. (\ref{eq:E-L equation}) to get the equations of motion for each mean field \cite{Zhu:2018qpv,Zhu:2018vwn}:
\begin{equation}
	\begin{aligned}
		m_{\sigma}^{2} \sigma+g_{2} \sigma^{2}+g_{3} \sigma^{3} &=-\frac{\partial M_{N}^{*}}{\partial \sigma} \rho_{s}, \\
		m_{\omega}^{ 2} \omega &=g_{\omega } \rho_{N}, \\
		m_{\rho}^{ 2} \vec{\rho} &=g_{\rho } \rho_{3},
	\end{aligned}
	\label{eq:field equations}
\end{equation}
where $\rho_{N} = \rho_{p}+\rho_{n}$, and $\rho_{3} = \rho_{p}-\rho_{n}$ that equals 0 in symmetric nuclear matter.

After the mean field approximation, Eq. (\ref{eq:field equations}) has become  algebraic equations, solving them and directly obtaining the information of the meson fields. To calculate the nuclear matter EOS, we should get the energy density and pressure from Lagrangian. They can be generated by Legendre transformation from Eq. (\ref{eq:lagrangian}) \cite{Zhu:2018qpv,Zhu:2018vwn}:
\begin{equation}
	\begin{aligned}
		\mathcal{E}=& \frac{1}{\pi^{2}} \sum_{i=n, p} \int_{0}^{p_{F}^{i}} \sqrt{p^{2}+M_{N}^{* 2}} p^{2} d p \\
		&+\frac{1}{2} m_{\sigma}^{2} \sigma^{2}+\frac{1}{3} g_{2} \sigma^{3}+\frac{1}{4} g_{3} \sigma^{4}-\frac{1}{2} m_{\omega}^{2} \omega^{2} \\
		&-\frac{1}{2} m_{\rho}^{2} \vec{\rho}^{2},
	\end{aligned}
	\label{eq:energy density}
\end{equation}
\begin{equation}
	\begin{aligned}
		P=& \frac{1}{3 \pi^{2}} \sum_{i=n, p} \int_{0}^{p_{F}^{i}} \frac{p^{4}}{\sqrt{p^{2}+M_{N}^{* 2}}} d p-\frac{1}{2} m_{\sigma}^{2} \sigma^{2} \\
		&-\frac{1}{3} g_{2} \sigma^{3}-\frac{1}{4} g_{3} \sigma^{4}+\frac{1}{2} m_{\omega}^{2} \omega^{2}+\frac{1}{2} m_{\rho}^{2} \vec{\rho}^{2}. 
	\end{aligned}
	\label{eq:pressure}
\end{equation}
From Eq. (\ref{eq:energy density}) and (\ref{eq:pressure}), the properties of nuclear matter can be determined.

In the earlier part of this section we discussed the standard method for calculating the EOS of nuclear matter. We also introduced the form that the nucleon number density should satisfy under the EVE model. If we consider the nonnegligible nucleon volume, then the modified nucleon density in EVE model should be used instead of the original Fermi gas model:
\begin{equation}
	p_F = (3\pi^2\rho^\prime)^{\frac{1}{3}},
	\label{eq:fermi-momenta}
\end{equation}
Eq. (\ref{eq:rhoscalar}) and (\ref{eq:rhoscalarINT}) should also be considered.

\subsection{Neutron Star Structure}
\label{subsec:NS-tov}
The nuclear matter EOS we considered only have the nucleons and mesons interactions. In order to get the NS matter, the leptons contributions must be considered to balance the positive charge of the proton and chemical potential of nuclear matter. The Lagrangian density of leptons can be written as
\begin{equation}
	\mathcal{L}_{\text {lep }}=\sum_{l=e, \mu} \bar{\Psi}_{l}\left(i \gamma_{\mu} \partial^{\mu}-m_{l}\right) \Psi_{l}.
	\label{eq:lepLagrangian}
\end{equation}
Since $\tau$ lepton has too large mass and decays easily, we only consider the first and second generation leptons -- $e,\ \mu$. The contributions from $e$ and $\mu$ for energy density and pressure are in the same way with Eq. (\ref{eq:energy density}, \ref{eq:pressure}):
\begin{equation}
	\begin{aligned}
		&\mathcal{E}_{\text {lep }}=\frac{1}{\pi^{2}} \sum_{l=e, \mu} \int_{0}^{p_{F}^{l}} \sqrt{p^{2}+m_{l}^{2}} p^{2} d p,\\
		&P_{\text {lep }}=\frac{1}{3 \pi^{2}} \sum_{l=e, \mu} \int_{0}^{p_{F}^{l}} \frac{p^{4}}{\sqrt{p^{2}+m_{l}^{2}}} d p.
	\end{aligned}
	\label{eq:leps-contributions}
\end{equation}
If we consider the conditions of $\beta$ equilibrium and charge neutrality, so we have
\begin{equation}
	\rho_{p}-\rho_{e}-\rho_{\mu}=0,
	\label{eq:charge}
\end{equation}
and
\begin{equation}
	\begin{aligned}
		\mu_{p}+\mu_{e} &=\mu_{n},\\
		\mu_{e}&=\mu_{\mu} ,
	\end{aligned}
	\label{eq:chemical}
\end{equation}
where \cite{Kouno:1995fs} 
\begin{equation}
	\mu_{i}=E_{F, i}^{*}+V_{i} P_{i}^{\prime}+g_{\omega} \omega\pm \frac{1}{2} g_{\rho} \vec{\rho}
	\label{eq:npchemical}
\end{equation}
is the chemical potential of nucleon with $V_p = V_n = 4\pi r_m^3/3$, ``$+$" for proton and ``$-$" for neutron. The function $P_{i}^{\prime}$ can be described as \cite{Kouno:1995fs,Mart:2013gfa,Suparti:2017msx}
\begin{equation}
	\begin{aligned}
		P_{i}^{\prime}=& \frac{1}{12 \pi^{2}}\left[E_{F}^{i} p_{F}^{i}\left(E_{F}^{i 2}-\frac{5}{2} M_{N}^{* 2}\right)\right.\\
		&\left.+\frac{3}{2} M_N^{* 4} \ln \left(\frac{p_{F}^{i}+E_{F}^{i *}}{M_N^{*}}\right)\right].
	\end{aligned}
	\label{eq:pfunction}
\end{equation}
Since leptons are point particles, their chemical potential can be written as
\begin{equation}
	\mu_{e(\mu)} = \sqrt{m_{e(\mu)}^2+p_F^{e(\mu)2}}.
	\label{eq:lepschemical}
\end{equation}

According Eq. (\ref{eq:charge}) to (\ref{eq:lepschemical}) the number density of nucleons and leptons could be obtained to generate the NS matter EOS.

To calculate the NS properties such like mass and radius of NS, we should consider the slowly rotating system must have spherical symmetry representation. Assuming that the
NS matter is a perfect fluid, one can obtain the Tolman-
Oppenhaimer-Volkoff (TOV) equation \cite{Tolman:1939jz,Oppenheimer:1939ne} which can describe the Einstein equation with slowly rotating NS metric \cite{Idrisy:2014qca},
\begin{equation}
	\begin{aligned}
		&\frac{d m}{d r}=4 \pi \mathcal{E} r^{2}, \\
		&\frac{d P}{d r}=-G \frac{\mathcal{E} m}{r^{2}}\left(1+\frac{P}{\mathcal{E}}\right)\left(1+\frac{4 \pi r^{3} P}{m}\right)\left(1-\frac{2 G m}{r}\right)^{-1},
	\end{aligned}
\label{eq:toveq}
\end{equation}
By substituting the EOS of the dense matter, i.e. $P(\mathcal{E})$, into the stellar structure equation (\ref{eq:toveq}), the mass-radius relationship of the NS and other properties can be obtained. Of course, we assume that the pressure at the NS radius $R$ is zero, which is $P(R) = 0$. Using this condition we can determine the mass for the radius $R$ of the NS \cite{Li:2020dst}.

\section{Results and discussions}
\label{sec:Results and discussions}
In the previous section, we have discussed the idea of calculating the EOS of NS, and presented the results of revising the mass radius of nucleons in the EVE model. It is important to note that since NS have the shell structures, we do not use the uniform EOS, although such works would be interesting. RMF method is used to calculate the core part of the NS, while BPS \cite{Baym:1971pw} and BBP \cite{Baym:1971ax} are used as inputs for the crust part. The density of nuclear matter at the critical point is $\rho_c = 0.09371$ fm$^{-3}$ \cite{Baym:1971pw,Baym:1971ax}.

\begin{figure}[H]
	\centering
	\includegraphics[width=0.45\textwidth]{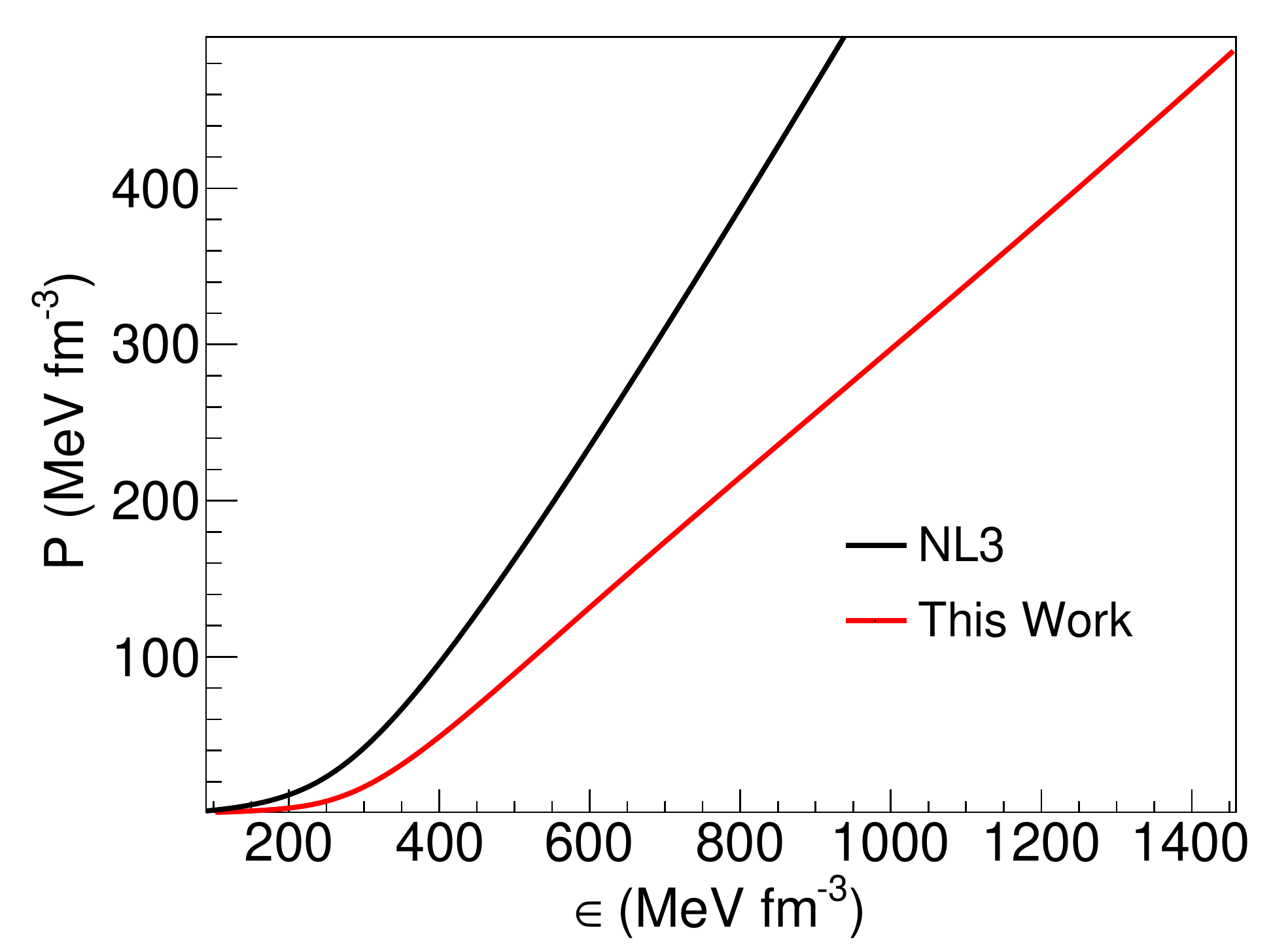}
	\caption{(Color Online) The EOS which is calculated by EVE correction with the mass radius of proton (neutron) and NL3 parametrization. The black solid line is about original NL3 and red is about this work.
	}
	\label{fig:EOS}
\end{figure}

\begin{figure}[H]
	\centering
	\includegraphics[width=0.45\textwidth]{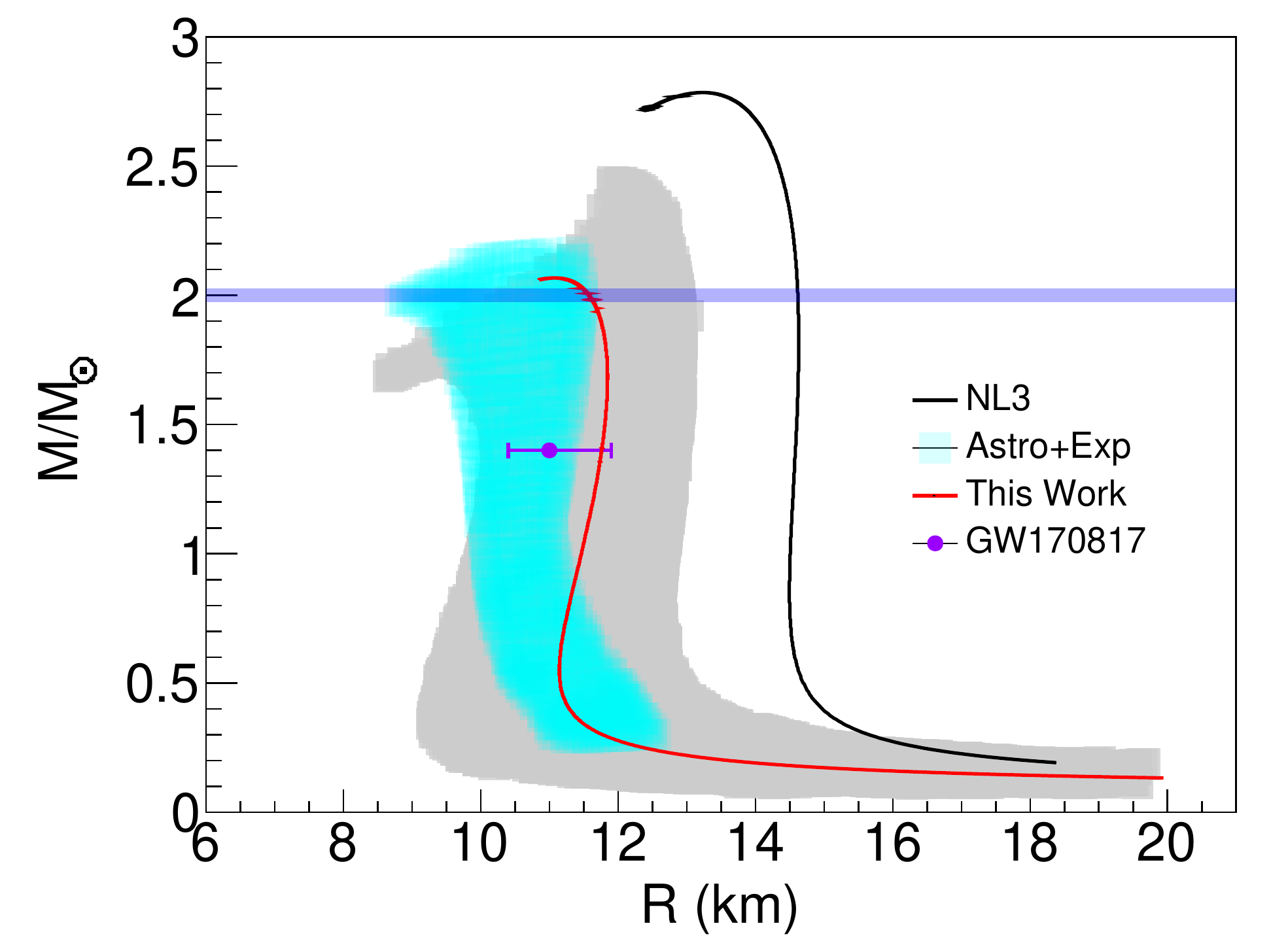}
	\caption{(Color Online) The mass-radius relations of NS. Horizontal
		shaded band in panel is  the pulsar mass constraint from
		Ref. \cite{Antoniadis:2013pzd}. Black solid line is about original NL3 and red is about this work. The constraint depicted as a cyan
		area was taken from Ref. \cite{Ozel:2016oaf}. Gray band comes from the uncertainty of the extraction of mass radius of proton \cite{Wang:2021dis}. Particularly, the purple one comes from GW170817 measurements \cite{Capano:2019eae}.
	}
	\label{fig:mass-radius}
\end{figure}

We follow Occam's razor, the only objective is to research the effect of the proton mass radius on the structure of NS. Parameters of the EOS of nuclear matter are consistent with the original NL3 sets \cite{Lalazissis:1996rd}. Our calculations and results are presented in the FIG. \ref{fig:EOS} and \ref{fig:mass-radius}.  

Figure \ref{fig:EOS} and \ref{fig:mass-radius} are our final results which calculated by EVE correction with mass radius of proton. From the results we find that our calculation is reasonable, and is consistent with GW170817 \cite{Capano:2019eae} after correction under the nucleon radius of $r_m = 0.55\pm0.09$ fm. Removing the EOS for the neutron star crust part, the two results are different but using the same RMF method and NL3 parameter sets (See FIG. \ref{fig:EOS}). The EOS we obtained looks a little bit ``Harder" than the original one. The ideal Fermi gas model tells us that nucleons are treated as point particles in an infinite nuclear matter moving thermally. But the components of nuclear matter inside a NS are complex. If the nucleon size is not negligible, the density of a finite-volume NS should be slightly smaller than the EOS of nuclear matter predicts. Based on previous work, we believe that the mass radius of proton is almost constant in free space, so they can be thought of as solid spheres packed tightly inside the NS. Compared with the ideal gas model, a NS of the same mass would have a smaller nucleon density. As the nucleon density decreases the pressure inside the NS decreases, too. Now we can go back to Eq. (\ref{eq:toveq}), when we solve it numerically we will get a smaller radius for the NS. So in FIG. \ref{fig:mass-radius} the radius we calculated by nucleon mass radius correction is smaller than the original NL3.

\section{Summary}
\label{sec:summary}
In this work, we applied the mass radius of nucleon extracted from our previous work and used it to calculate the mass and radius information of the NS. According the results we obtained, the size of the nucleons in NS must be considered. On one hand, we believe that EVE model is feasible to modify the EOS of nuclear matter of NS. But the premise is that the mass radius of the nucleon is used instead of the charge one. On the other hand, the calculation result of the radius of NS which has the 1.4 times solar-mass is consistent with the experiment--GW170817 \cite{Capano:2019eae}. The gray error band in FIG. \ref{fig:mass-radius} of our calculation results depends on the uncertainty of the extracted nucleon mass radius. At present, it seems that the error is relatively large. In the future, more precision experimental data from Electron-ion Collider (EIC) \cite{Accardi:2012qut,AbdulKhalek:2021gbh} for US and Electron-ion Collider in China (EicC) \cite{Chen:2018wyz,Chen:2020ijn,Anderle:2021wcy} will reduce the uncertainty of the results.

\begin{acknowledgments}
We are very grateful to Profs. Jianmin DONG and Ang LI for their fruitful suggestions and the discussions.
This work is supported by the Strategic Priority Research Program of Chinese Academy of Sciences
under the Grant NO. XDB34030301.
\end{acknowledgments}

\bibliographystyle{apsrev4-1}
\bibliography{refs}

\end{document}